\documentclass[a4paper,pra,twocolumn,floatfix]{revtex4-1}  

\usepackage{graphicx}
\usepackage{epsfig}
\usepackage{latexsym}

\begin{document}



\title{Melting temperature of graphene}
\author{J. H. Los, K.V. Zakharchenko, M. I. Katsnelson, Annalisa Fasolino*}

\affiliation{Radboud University Nijmegen/Institute for Molecules and Materials,
                Heyendaalseweg 135, 6525AJ Nijmegen, The Netherlands}
\affiliation{* a.fasolino@science.ru.nl}\date{\today}

\begin{abstract}
We present an approach to the melting of graphene based on nucleation 
theory for a first order phase transition from the 2D solid to the 3D liquid via an intermediate quasi-2D liquid. 
 The applicability of nucleation 
theory, supported by the results of systematic atomistic Monte Carlo 
simulations, provides an intrinsic definition of the melting temperature 
of graphene, $ T_m $, and allows us to determine it. We find $T_m \simeq 4510$ K, 
about 250 K higher than that of graphite using the same interatomic
interaction model. The found melting temperature is shown to be in good 
agreement with the asymptotic results of melting simulations for 
finite disks and ribbons of graphene. Our results strongly 
suggest that graphene is the most refractory of all known materials.
\end{abstract}

\maketitle 
\section{Introduction}

Surprisingly, understanding of melting is still an open problem in 
condensed matter physics. For instance, after more than hundred years, 
there is still no reliable theoretical justification of the Lindemann 
criterion \cite{bunch}. Nevertheless, the melting temperature $ T_m $ 
is well-defined as the temperature at which the Gibbs free energy curves
of the liquid and solid phase intersect. Graphene, however, is a 
quasi-2D crystalline membrane \cite{ACR2013}, which melts into a 
3D liquid phase \cite{Zakharchenko}. The different dimensionality of these 
two phases makes it impossible to determine their free energy difference with 
existing methods, so that it is a priori not even clear how to define $ T_m $.
Here we show that a 2D nucleation theory offers a way to overcome this difficulty 
and give a reliable quantitative value of $ T_m $.

While for 3D systems, there is no consensus about how to describe 
melting and premelting anomalies, for a 2D crystal there is a 
commonly accepted microscopic scenario for melting \cite{Gasser}, 
the KTHNY theory \cite{KT,HN,NH,Y,Chen}. In the KTHNY theory, melting 
occurs via unbinding of topological defects, like vortices in superconductors
\cite{Larkin}. This scenario seems not to apply to graphene.
Apart from the embedding in 3D space, which affects all its 
structural and thermodynamic properties \cite{ACR2013,Fasolino},
the nature of defects is completely different from those considered 
in the KTHNY theory. 
In graphene, there are neither observations nor predictions of stable 
single pentagons or heptagon defects (disclinations). Dislocations 
(pentagon-heptagon pairs) have been observed mainly at grain boundaries 
\cite{Coraux}. The 4.6 eV formation energy \cite{LCBOPII} of a 
Stone-Wales (SW) defect made of two adjacent  pentagon-heptagons (57) pairs 
is much lower than the formation energy of two well separated 57 pairs 
as occurring in small angle grain boundaries \cite{Carlsson}.
In the Appendix we show how the formation 
energy of two 57 pairs rapidly increases with separation.
The relative stability of SW defects make them play a crucial role 
in the pre-melting behavior of graphene \cite{Zakharchenko} and
prevents them to split into two 57 defects and further, 
and by this to follow the KTHNY scenario. 

In ref.\onlinecite{Zakharchenko} we have shown that, within the same model of interactions that we use here, spontaneous melting of graphene occurs at  a temperature that we call $T_m^*$ of 4900  K, giving an upper limit 
 for the true melting temperature $T_m$ of graphene.
Here we show that the melting of bulk graphene can be described by
nucleation theory, allowing an unambigous definition of the melting temperature, $ T_m $.
We find $T_m \simeq 4510$ K, 
about 250 K higher than that of graphite\cite{Colonna} and sofar the highest of all materials.

The paper is organized as follows.
In section II we review previous results \cite{Zakharchenko} on the spontaneous melting of bulk graphene. 
In section III we introduce an approach based on classical and kinetic nucleation theory to the melting in 2D.
This approach is applied in section IV to bulk graphene. The melting of finite disks and ribbons is presented in sections V and VI respectively. Summary and conclusions are given in section VII. 

\section{Spontaneous melting}

In a previous work \cite{Zakharchenko}, we have studied the spontaneous 
melting of graphene by means of Monte Carlo (MC) simulations based 
on the reactive bond order potential LCBOPII \cite{LCBOPII}. 
A suitable Lindemann type order parameter for graphene was defined as
$ \gamma_{12}=(1/a^2) < ( {\bf r}_i - (1/12) \sum_j {\bf r}_j )^2 > $,
where $ a = 1/\sqrt{\pi \rho} $ is the atomic radius, with $ \rho $ the 
2D particle density, $ {\bf r}_i $ is the position of the $i$-th 
atom and the sum over $j$ runs over the $12$ atoms closest to atom $i$. 
For graphene, melting starts at $\gamma_{12} \simeq 0.1$, close to the 
value found for a strictly 2D, triangular lattice \cite{Bedanov} . 

\begin{figure}[htb]
\hspace*{-3.3cm}a

\vspace*{-0.25cm}
\includegraphics[width=3.cm,clip]{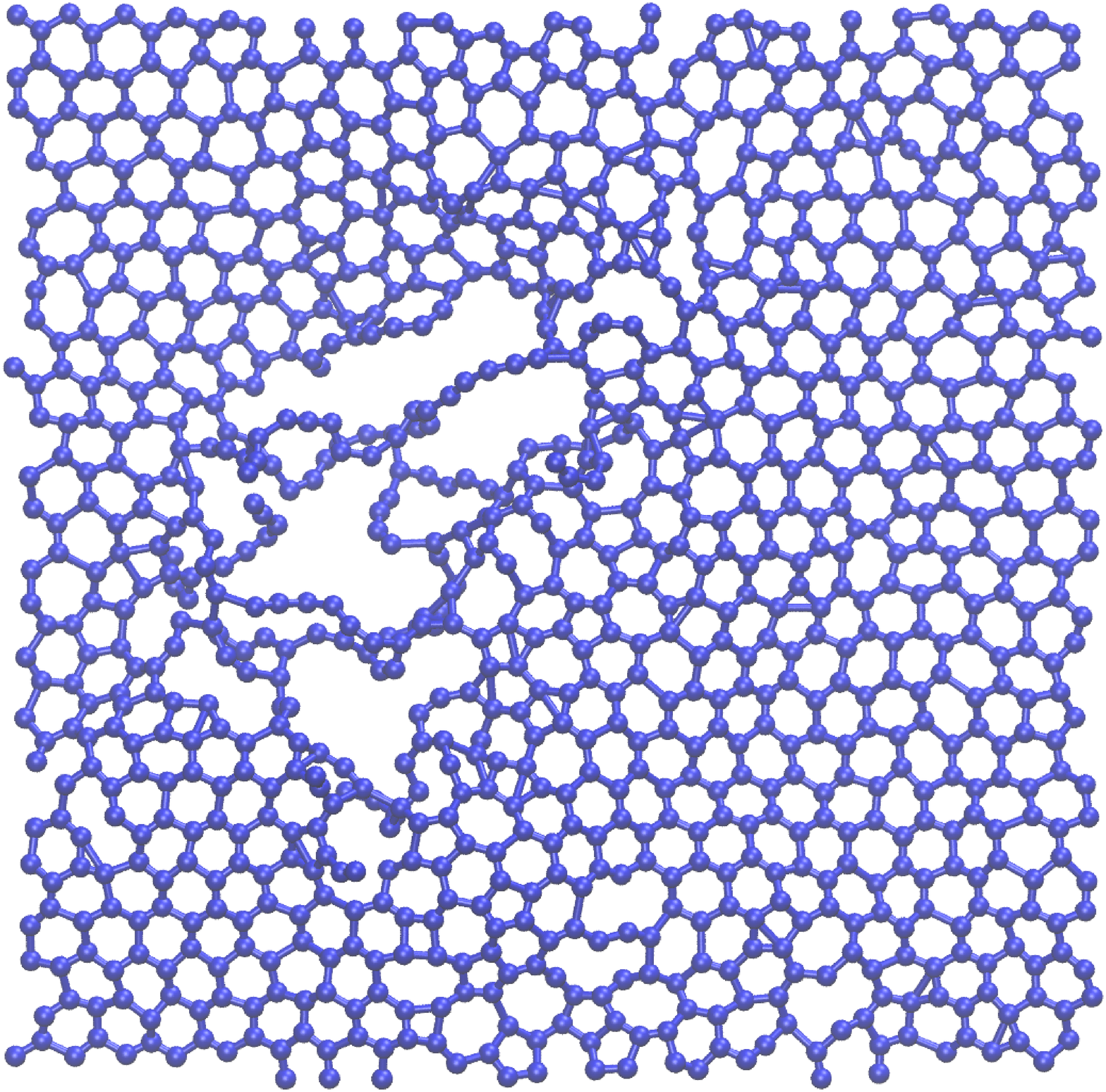}
\includegraphics[width=.45\textwidth,clip]{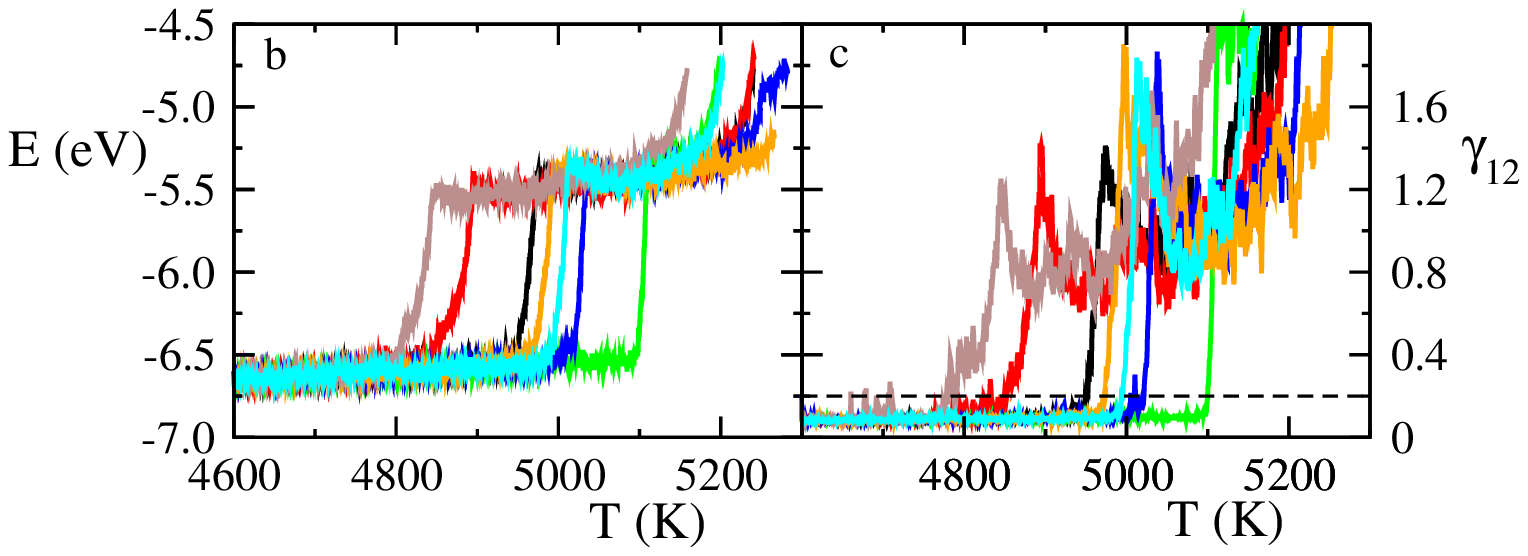}
\caption{ \label{Egamma}
a) A snapshot of simulated bulk graphene at the onset of melting. 
The system contains 1008 atoms with PBCs.
Note the coexistence of solid and liquid parts.
b) Energy $E$ and c) and order parameter $ \gamma_{12} $ 
as a function of temperature $T$ during a linear temperature ramp
for 8 independent simulations. }
\end{figure}

In Fig. \ref{Egamma}a we show a snapshot from a MC simulation 
in which a bulk graphene system consisting of $N=$1008 atoms with 
periodic boundary conditions (PBCs) is heated at a constant rate.
Fig. \ref{Egamma}b and c show the potential energy $E$ and $\gamma_{12}$ 
as a function of the applied, increasing temperature found in 
8 independent MC simulations. Melting is signalled by a jump in 
$E$ and $\gamma_{12}$, but with large variations in the observed 
melting temperature, $T_m^* $, at which this jump occurs for
different, independent simulations. This jump as well as the coexistence 
of solid and liquid parts in Fig. \ref{Egamma}a are typical features 
of first order phase transitions. Obviously, the lowest $T_m^*=4900 $ K 
found in Ref. \onlinecite{Zakharchenko} constitutes only an 
upper limit for the true melting temperature $T_m$ of graphene. 

\section{Nucleation theory approach to melting in 2D}

Here we propose a quantitative approach to determine 
$ T_m $ by considering the melting as a process initiated and dominated 
by the nucleation of liquid nuclei in the solid sheet, a typical scenario 
for a first order phase transition.
The possible applicability of nucleation theory is furthermore suggested
by our observation that the melting process occurs in two steps.
First the graphene sheet transforms into a sort of quasi-2D liquid
phase consisting of entangled and interconnected chains remaining 
roughly within the quasi-2D plane of the previously solid sheet.
Eventually, the chains start to disentangle and
extend to 3D space, with a diverging simulation box in NPT
simulation at zero pressure applied in this work.
In Fig. \ref{Egamma} this "graphene $ \rightarrow $ 
quasi-2D liquid $ \rightarrow $ 3D liquid" scenario is reflected in the
energy curve: first, at melting, the energy steeply increases by about 
1.05 eV after which it remains roughly constant for a while before it
starts to increase further. This two steps scenario via 
an intermediate quasi-2D liquid phase
would imply that the melting temperature of graphene is in fact the 
temperature at which the 2D solid and the quasi-2D liquid phases are in 
equilibrium. 
Looking at the snapshot in Fig. \ref{Egamma}a, it can also be argued that 
before complete melting, when liquid nuclei start to form in the solid sheet, 
the liquid phase remains connected with the solid sheet at the edges 
of the nucleus, constraining it to stay within quasi-2D configurations. 
The importance of assuming the existence of a quasi-2D liquid phase with 
its own thermodynamic properties is that it allows us to formulate 
a nucleation theory for the melting of a 2D solid embedded in 3D space, 
the case of graphene. If the melting can only take place via this 
intermediate, quasi-2D liquid phase, then this intermediate phase is 
decisive for the location of the melting temperature of graphene. 

For the analysis of melting in terms of nucleation theory 
we performed two series of simulations: simulations at constant 
temperature and simulations applying a linear temperature ramp.  
To assess the 
accuracy of our approach, we will compare classical nucleation theory 
(CNT) and kinetic nucleation theory (KNT), as described hereafter.
We use LCBOPII \cite{LCBOPII} for the interatomic interactions, 
as in Ref. \cite{Zakharchenko}. We point out that the melting 
temperature of graphite according to LCBOPII has been accurately 
determined to be 4250 K \cite{Colonna}. 

For the application of CNT to melting in  2D, we write the work 
$ W(r) $ done to form a quasi-2D liquid nucleus of radius $r$ as:
\begin{equation} 
\label{Wr}
W (r) = \pi r^2 \rho \Delta\mu + 2\pi r \gamma_{sl}
\end{equation}
where $\Delta\mu=\mu_l-\mu_s$ is the chemical potential difference
per particle between the quasi-2D liquid phase and the solid phase,
and $ \gamma_{sl} $ is the solid-liquid interface free energy. 
Note that $ \pi r^2 \rho $ is the number of liquid particles 
inside the nucleus, which should indeed grow as $ r^2 $ due to the 
2D geometry of the solid. For $T>T_m$, $\Delta\mu < 0$ and $ W (r) $ 
has a maximum at the critical nucleus size with radius 
$ r_c = - \gamma_{sl}/( \rho \Delta \mu ) $
equal to 
$ W_c \equiv W (r_c) = - \pi \gamma_{sl}^2/( \rho \Delta \mu ) $. 
Using the thermodynamic relation
$ \Delta\mu=\Delta h -T \Delta s \approx (\Delta h/T_m) \left (T_m-T \right) $,
with $ \Delta h $ and $ \Delta s = \Delta h/T_m $ the melting enthalpy 
(latent heat) and melting entropy per particle respectively, the 
nucleation probability is derived to be:
\begin{equation}
\label{Pnucl} 
P^{CNT}_{nucl} \equiv K \exp{\left(- \beta W_c \right) } = 
K \exp{\left( - \frac{\beta~ \alpha T_m}{ (T-T_m) }\right)} 
\end{equation}
where $ K $ is a kinetic prefactor, $ \beta = 1/(k_B T) $ and we defined
$ \alpha \equiv \pi \gamma_{sl}^2 / ( \rho \Delta h) $. 

If the melting is dominated by nucleation, then the average time $ t^* $ 
required for melting to occur is given by the solution of the equation:
\begin{equation}
\label{Nnucl}
N_{nucl}(t^*) = \int_0^{t^*} P^{CNT}_{nucl} (T(t)) dt = 1 
\end{equation}
where $ N_{nucl} (t) $ is the number of (super)critical nuclei at time $t$.
At constant temperature $ T > T_m $, eq. \ref{Nnucl} simplifies to 
$ t^* = 1/P_{nucl}(T) $. Instead, if the system is slowly heated 
at a heating rate $ \eta $, such that $ T = T_m + \eta t $, we have 
to solve eq. \ref{Nnucl} numerically for $t^* $.
The temperature at which melting is expected, $ T^*_m $, 
is then readily obtained from $ T_m^* = T_m + \eta t^* $.

In principle, CNT does not specify neither the prefactor $ K $ nor its 
temperature dependence. Therefore, we take $K$ constant in our simulation 
analysis based on CNT. Nucleation is then determined by three
parameters: $ \alpha $, $K$ and $T_m$.
Expressions for the temperature dependence of the
kinetic prefactor belong to the domain of kinetic nucleation theory (KNT), 
which allows us to perform a more accurate analysis.

The starting point of KNT \cite{Kashchiev} is quite different
from that of CNT. KNT assumes a (liquid) cluster size distribution 
(CSD) governed by a coupled set of master equations with 
appropriate boundary conditions. In CNT, the CSD is proportional to 
$ \exp{(-\beta W(r))} $ so that it has a minimum for clusters of the critical 
nucleus radius $ r_c $, where $W$ is maximal, while it increases exponentially 
beyond the critical size. The latter prediction is clearly unphysical.
In KNT, this unphysical behavior is avoided by requiring the CSD to satisfy 
the boundary condition $ X_N = 0 $, where $ X_N $ is the number of clusters 
that incorporate all $N$ atoms of the system. Eventually, KNT leads to 
a stationary state nucleation rate given by:
\begin{equation}
\label{Jnucl} 
P^{KNT}_{nucl} = f_c Z \exp{\left( - \frac{\beta ~ \alpha T_m}{ (T-T_m) }\right)}
\end{equation}
where $ f_c $ is the attachment rate for a particle to join a nucleus
of the critical size, $ Z $ is the so-called Zeldovich factor and $ \alpha $
as defined before.
The attachment rate can be expressed as:
\begin{equation}
\label{fc} 
f_c = \nu_{ref} \frac{T}{T_{ref}} g \sqrt{n_c} 
\exp{ \left( -\beta E_a \right) }
\end{equation}
where $ \nu_{ref} \simeq k_B T_{ref}/h $ (with $h$ the Planck constant)
is an attempt frequency at some reference temperature $ T_{ref} $,
which we choose to be $ T_{ref} = 4500 $ K, $ g \sqrt{n_c} $ 
accounts for the number of perimeter sites around a critical nucleus
with $ n_c = \pi r_c^2 \rho = \pi \gamma_{sl}^2/(\rho \Delta \mu^2) $ and
$g$ a geometrical factor, while $ E_a $ is an activation energy
barrier for the attachment process. For a circular nucleus, 
the number of perimeter sites is given by 
$ 2 \pi r_c d \rho = (2 d/a) \sqrt{n_c} $ where $ d $ is the width 
of one atomic layer (or shell), whence $ g=2d/a $.
Following Ref. \cite{Kashchiev}, the Zeldovich factor $Z$ for our
2D case is found to be:
\begin{equation}
\label{Zeld} 
Z = 
\left( \frac{ \beta }{ 4 \pi } \frac{ | \Delta \mu |^3 }{ \alpha \Delta h} \right)^{1/2}.
\end{equation}
Comparing the parameters in KNT to those in CNT, and noting 
that $ \tilde{K} \equiv \nu_{ref} g $ replaces $K$ in CNT, 
there are two additional parameters: $ \Delta h $ and $ E_a $.
However, $ \Delta h $ can directly be obtained from the simulations as the 
melting energy (at zero pressure): $ \Delta h \simeq 1.05 $ eV (see Fig. \ref{Egamma}b).
The activation energy barrier $ E_a $ is typically a few eV, implying that
the temperature dependence introduced by $ \exp{(-\beta E_a)} $ is rather weak within 
the temperature range of interest and has only a very minor effect (as we verified).
We have taken $ E_a $ = 2.6 eV fixed, equal to the activation energy for 
diffusion in a mixed sp-sp$^2$ carbon liquid, as found with LCBOPII \cite{Colonna}.

\begin{figure}[htb]
\includegraphics[width=.3\textwidth,clip]{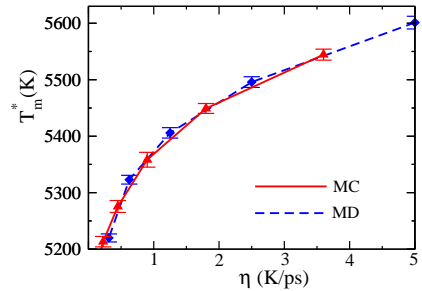}
\caption{ \label{MC-MD-noMR}
Comparison of the average, observed melting temperature, $ T_{m}^*$, of 
bulk graphene (1008 atoms with PBCs) as a function of the heating rate
obtained from MD (blue diamonds) and MC (red triangles) simulations 
after calibration, using LCBOPII without the middle range interactions. 
Error bars are based on 48 independent simulations. Lines are 
guides to the eye.}
\end{figure}

LCBOPII include short range (covalent), long range (van der Waals) 
and middle range (MR) interactions \cite{LCBOPII}.  The latter 
contribution is a correction that improves the reactive properties and, 
by construction, does not affect the equilibrium carbon structures at low $T$.
Since many simulations were required to obtain sufficient statistics, and the use 
of the full potential including the MR part is computationally demanding for 
Molecular Dynamics (MD) simulations, we have performed  MC simulations.
Following Ref. \onlinecite{Huitema}, we then assume that time is proportional
to the number of MC cycles $t_{MC}$ (one cycle being $N$ trial moves) so that
$ t = C t_{MC} $ where $C$ is a time calibration factor assumed to be constant 
within a limited temperature range provided that the acceptance rate is kept constant.
To assess the validity and accuracy of this approximation,
we also performed Molecular Dynamics simulations, but 
without the MR part of the LCBOPII for the sake of feasibility, and compared 
the results to those from MC simulations using the same potential (i.e.
LCBOPII without the MR part). The MC and MD simulations were performed 
applying a linear temperature ramp.
In the MD simulations the temperature was raised by using the 
Berendson thermostat and scaling the velocities at every time step, 
while in the MC simulations the temperature was raised every 5 MC cycles. 
We considered a series of heating rates, and for each of them 
the average $T_{m}^*$ was determined from 48 independent 
simulations. For each simulation, $ T^*_m $ was determined by the
intersection of the $ \gamma_{12} $ curve with the horizontal 
line at $\gamma_{12} $ = 0.2 (see Fig. \ref{Egamma}a).
Plotting the average $T_{m}^*$ as a function of $ \eta $, a very good 
agreement between the MD and MC results was obtained by applying a time 
calibration factor equal to $ C=5.56~10^{-4} $ ps/cycle, as is shown 
in Fig. \ref{MC-MD-noMR}. We note that for the determination of $ T_m $ 
using nucleation theory, it is sufficient that $ t=C t_{MC} $ holds and 
we do not need $C$.

\begin{figure}[htb]
\includegraphics[width=.45\textwidth,clip]{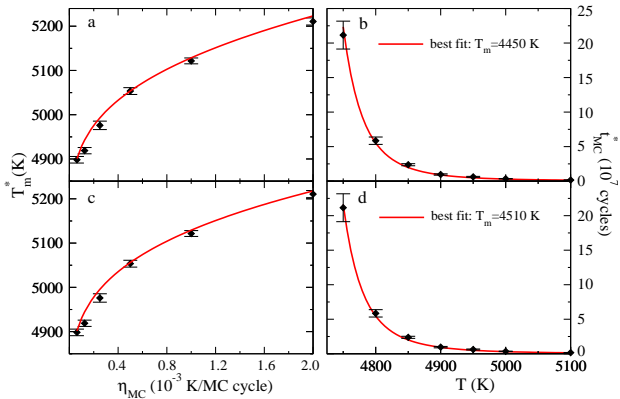}
\caption{\label{TmFromNT}
Symbols: a) and c) average melting temperature $ T_m^* $ as a function of the
MC heating rate $ \eta_{MC} $ and b) and d) MC 'time' $ t^*_{MC} $ required 
for melting at isothermal conditions as a function of temperature.
The error bars are based on at least 48 independent simulations, but those for 
the isothermal simulations at 4750, 4800 and 4850 K are based on 96 simulations.
The solid (red) curves represent simultaneous best fits of both sets of 
simulation data based on predictions by CNT (a and b) and KNT (c and d).}
\end{figure}

\section{Melting of bulk graphene}

We apply the CNT and KNT described in the previous section to evaluate $T_m$ of  a bulk graphene system consisting of $N=$1008 atoms with PBCs as in Fig. \ref{Egamma}.
In Fig. \ref{TmFromNT} we present the results of two sets of systematic 
MC simulations, including simulations heating the system at a rate $ \eta_{MC} $ 
and isothermal simulations. In view of the large variations in the
respective, observed melting temperature $ T_m^* $ and time required 
for melting, $ t^*_{MC} $, at least 48 independent simulations were 
performed in all cases, in order to obtain good statistical averages. 
Best fits to all 13 data points were then determined by using
the described CNT and KNT approaches, with fitting parameters
$K$ (or $\tilde{K}= \nu_{ref} g $ in the case of KNT), $ \alpha $ and $ T_m $. 
In both cases good agreement was obtained, with estimates for $ T_m $ 
equal to 4450 and 4510 K according to CNT and KNT respectively. This relatively
small difference shows that the exponential factor dominates the nucleation 
rate and that the role of the temperature dependence of the prefactor 
in KNT is limited. Since KNT is more accurate,
we conclude that $ T_m \simeq 4510 $ K.
From $ \alpha $ we find $ \gamma_{sl} = 0.134$ eV/\AA~.

\begin{figure}[htb]
\hspace*{-3.5cm} a \hspace*{3.0cm} b

\vspace*{-0.5cm}
\hspace*{-0.18cm} \includegraphics[width=3.cm,clip]{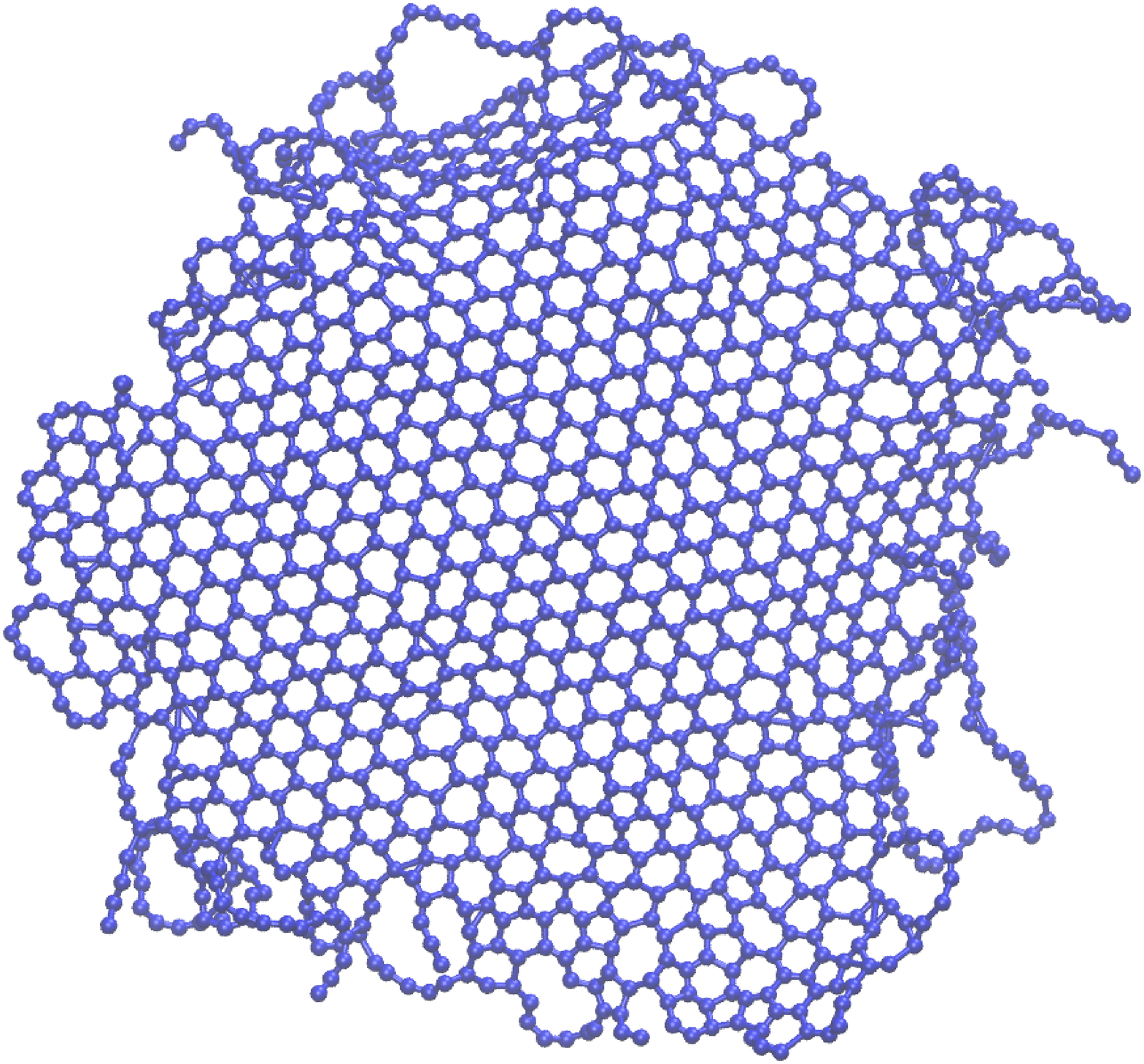}
\hspace*{0.2cm} \includegraphics[width=3.cm,clip]{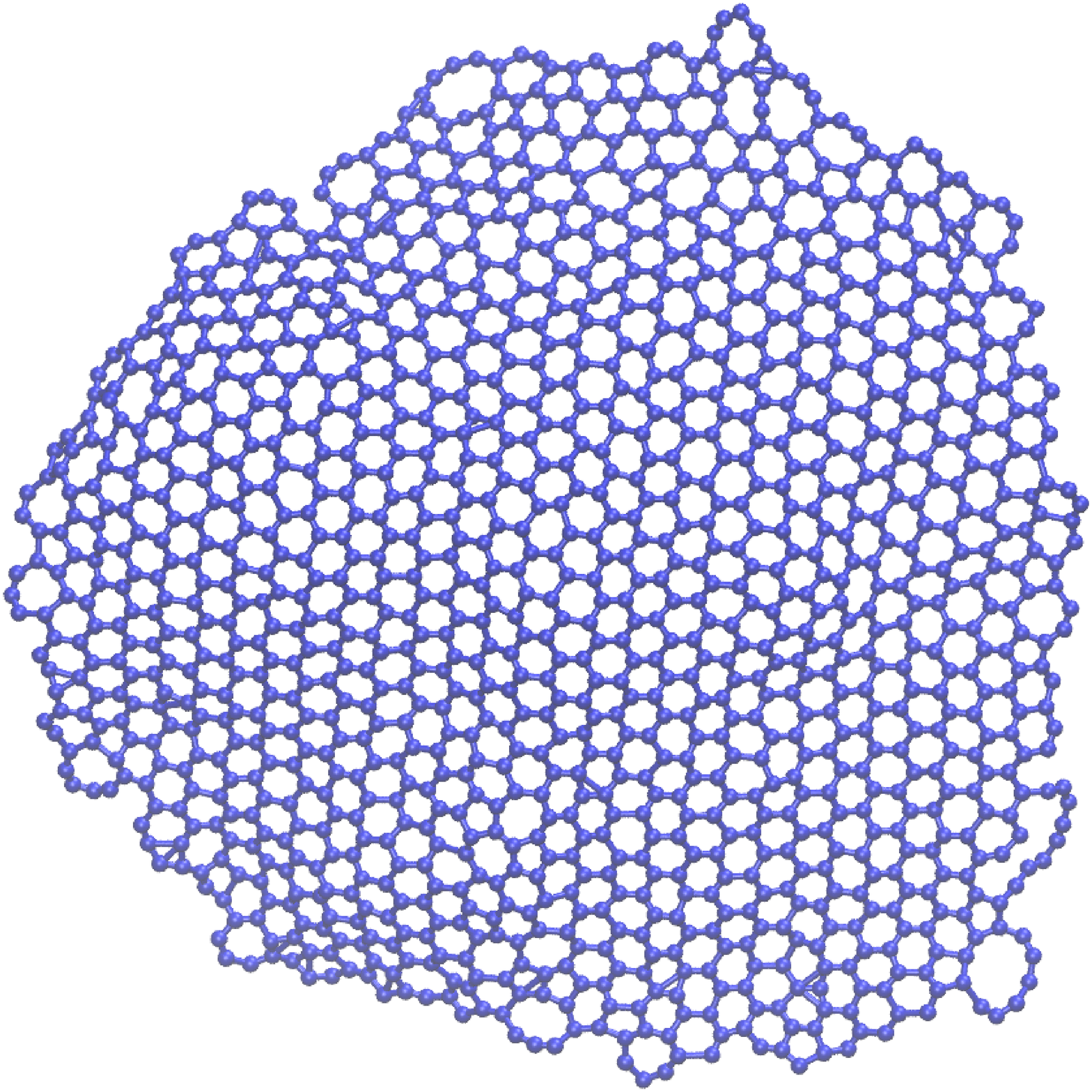}

\includegraphics[width=.45\textwidth,clip]{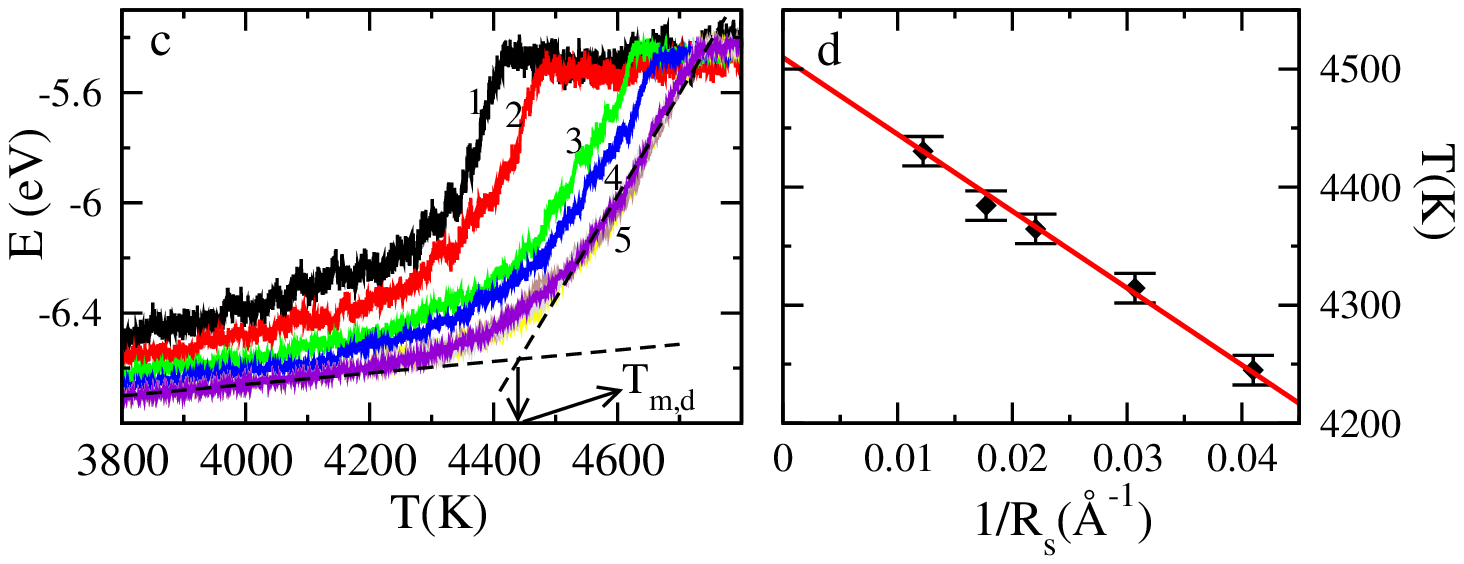}
\caption{\label{TmFromTmCl}
a) Snapshot at 4300 K, illustrating edge melting of a graphene disk 
containing 1274 atoms heated at a rate of 2.5 10$^{-4}$ K/cycle.
b) Snapshot illustrating the recrystallization 
after cooling the system in a) from 4300 to 3800 K.
c) Melting curves for different graphene disk sizes:
1) 714, 2) 1274, 3) 2470, 4) 3820 and 5) 8020 atoms. For the latter
cluster, melting curves from 4 independent simulations are drawn but 
they are almost overlapping. The crossing dashed lines illustrate the 
method used to define the melting temperature $ T_{m,d} $ of a disk. 
d) Best fit (red line) according to eq. \ref{Tmd} of the 
average simulated disk melting temperatures (symbols). The error
bars are based on 4 independent simulations.}
\end{figure}

\section{Melting of finite graphene disks}

Searching for verfication of the result from KNT, we also performed simulations 
of the melting of a series of roughly circular graphene disks of increasing size.
Typically, their melting temperature, $ T_{m,d} $, increases as a function 
of cluster size, as was already demonstrated for small graphene flakes 
($ N < 55$) in Ref. \onlinecite{Singh}. According to the Pavlov model~\cite{Pavlov} 
specified below, $ T_{m,d} $ deviates from $ T_m $ by a finite size correction that
vanishes with the inverse radius of the disk, allowing for the determination 
of $ T_m $ by extrapolation. It should be noticed, however, that the analysis 
of the disk simulations is arguable because 
$ T_{m,d} $ is not rigorously defined and  the Pavlov model does not 
consider edge melting.

For a disk, nucleation is not 
required and the melting starts at the edge, as illustrated in 
Fig. \ref{TmFromTmCl}a. According to CNT there is 
no barrier for nucleation at a 1D edge, which is rough 
at any $T$. Therefore there is little variation in $ T_{m,d} $ from 
different simulations, as is illustrated by the almost overlapping melting
curves from 4 independent simulations for the cluster with 8020 atoms  
in Fig \ref{TmFromTmCl}c. 

Interestingly, the melting process is reversible to some extent, 
as long as there are still solid parts in the system.
In Fig. \ref{TmFromTmCl}b we show the 
recrystallization that has occurred after the system in 
Fig. \ref{TmFromTmCl}a was slowly cooled from 4300 to 3800 K.

An appropriate approach to analyse our simulation 
data for disks would be a 2D version of Pavlov's model
\cite{Pavlov}. Writing the free energy of the solid ($s$) and quasi-2D liquid ($l$) 
disk as $ G_{d,P} = N_P \mu_P + 2 \pi R_P \gamma_{Pv} $ for $ P=s,l $,
with $ N_P $ the number of particles in the solid/liquid disk, $ R_P $ their 
radii and $ \gamma_{Pv} $ the respective edge free energies of their interfaces 
with the vacuum, and imposing equality of the chemical potentials through 
$ \left . dG_{d,s}/dN_s \right|_{N_s=N} = \left . dG_{d,l}/dN_l \right|_{N_l=N} $ 
leads to the following correction for the melting temperature of a finite disk:
\begin{eqnarray} 
\label{Tmd}
T_{m,d} = T_m 
\left( 1 - \frac{ \gamma_{sv} - q_{\rho} \gamma_{lv} }{\rho_s \Delta h R_s} \right)
\end{eqnarray}
where $ q_{\rho} \equiv (\rho_s/\rho_l)^{1/2} $ accounts for the 
2D density difference between solid and liquid.

The results of the disk simulations are summarized in Fig. \ref{TmFromTmCl}.
A clear trend in $ T_{m,d} $  as a function of $ 1/R_s $
is observed and a best fit based on eq. \ref{Tmd} yields $ T_m = 4503 $ K, 
a value very close to that derived from KNT, and 
$ \gamma_{sv} - q_{\rho} \gamma_{lv} = $ 0.58 eV/\AA~. 
With an (average) value of $ \gamma_{sv} \simeq $ 1.05 eV/\AA~ for a 
graphene ribbon from Ref. \cite{Kroes} and taking $ q_{\rho} \simeq 1 $
we find $ \gamma_{lv} \simeq 0.47 $ eV/\AA~. These edge energies, and also
$ \gamma_{sl} = 0.134$ eV/\AA~ from KNT reported above, are physically 
sound, with $ \gamma_{sv} > \gamma_{lv} > \gamma_{sl} $.

\section{Melting of 2D graphene ribbons}

\begin{figure}[htb]
\hspace*{-5.50cm}a \hspace*{3.cm}b

\vspace*{-0.5cm}
\includegraphics[width=2.7cm,clip]{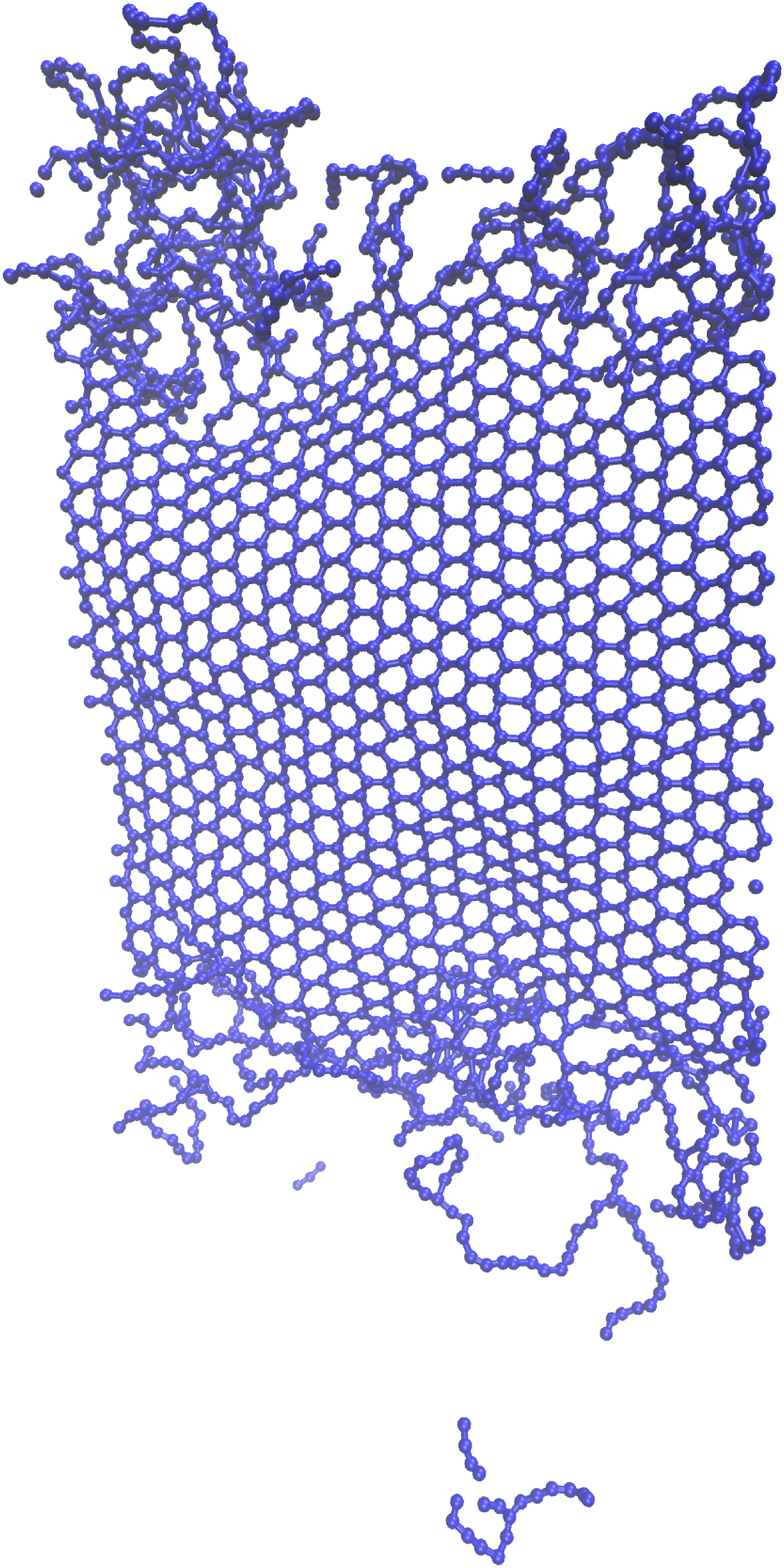}
\includegraphics[width=.315\textwidth,clip]{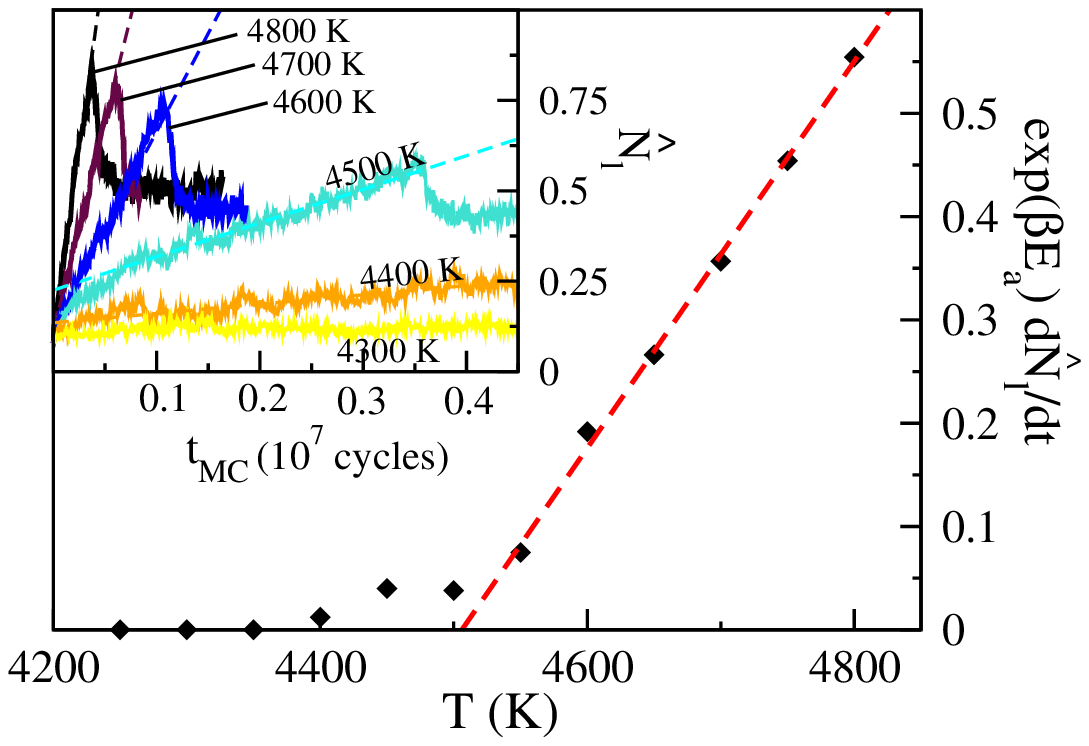}
\caption{\label{ribMlt}
 a) Snapshot of a melting  ribbon consisting of 2016 atom at $T$ = 4500 K.
b) The simulated melting speed (symbols), $\exp{(\beta E_a) d\hat{N}_l/dt} $, as 
a function of the temperature $T$ for a graphene ribbon. The inset shows 
the fraction of liquid particles $ \hat{N}_l $ as a function of $ t_{MC} $ for 
various simulation temperatures as indicated. For the completely liquid phase 
$ \hat{N}_l \simeq 0.5 $ as many particle still have a $ \gamma_{12} < 0.2 $ 
due to a rather isotropic neighbor surrounding.}
\end{figure}

For further verification of our result for $ T_m $, we have also 
performed constant temperature melting simulations of a graphene 
ribbon. We used a ribbon of a size $ L_x \times L_y $ = 51.12 $ \times $ 103.23
\AA$^2$ containing 2016 atoms with PBC only in the 
x-direction. For this geometry, as for the disks, the melting proceeds from 
the free edges, as shown in Fig. \ref{ribMlt}a, and there is no nucleation 
barrier. 
In this case of a straight edge, the melting speed, $ dN_l/dt $ 
with $ N_l $ the number of liquid particles, which we define as the number
of particles with $ \gamma_{12} > 0.2 $, is expected to behave as 
$ \hat{K}(T) P_{sl} ( \exp{( - \beta \Delta \mu ) } - 1 ) $, where
$ P_{sl} \simeq 2 L_x d/(\pi a^2) $ is the number of sites at the 
solid-liquid interface and $ \hat{K}(T) = (\nu_{ref} T/T_{ref}) \exp{(- \beta E_a ) } $. 
Since $ | \beta \Delta \mu | << 1 $, the melting speed can be well approximated as:
\begin{equation} 
\label{dNldt}
\frac{d\hat{N}_l}{dt}= 
\frac{\nu_{ref} \Delta h \exp{( - \beta E_a )} }{k_B T_{ref} T_m} 
\frac{2 L_y d \rho}{N} ~ ( T - T_m )
\end{equation}
where we have defined $ \hat{N}_l \equiv N_l/N $ as the fraction of liquid particles.
In Fig.  \ref{ribMlt}b we have plotted $ \exp{(\beta E_a) d\hat{N}_l/dt } $ 
as a function of $ T $ obtained from simulations (symbols). The intersection 
of a best linear fit according to Eq. \ref{dNldt} for $ T > 4500 $ K with the 
horizontal axis yields $ T_m = $ 4505 K, very close to the previously found values. 
It has to be noticed, however, that the results for $ T = $ 4400, 4450 and 4500 
do not follow the linear law, as the system actually melts for these temperatures.
Although finite size effects might play a role here, maybe the most important
reason is that chain-like molecular units evaporate from 
the liquid edges, as illustrated in Fig. \ref{ribMlt}a.
Since equilibrium in the present 
case is characterized by a liquid edge of fixed average width, evaporation 
must lead to a thinner and thinner solid core and finally to complete melting. 
To deal with this issue properly would require to consider a three phase equilibrium. 
From the simulation results for $ T > $ 4500 K, using Eq. \ref{dNldt} 
and the previously found time calibration factor $ t = 5.56~ 10^{-4} $ ps/cycle,
one can directly determine the attempt frequency $ \nu_{ref} $. 
We find $ \nu_{ref} = 2.3~10^{14} $ Hz, which is quite reasonable
when comparing it to $ k_B T_{ref}/h = 0.94~10^{14} $ Hz. 

\section{Summary and conclusions}

By analyzing systematic simulations we have shown that the 
melting of pristine bulk graphene, a prototype 2D solid
embedded in 3D space, follows a two-stage scenario 
of which the first step from the 2D solid to a quasi-2D liquid 
phase is well described by nucleation theory for first order 
phase transitions. As a consequence, graphene has an unambiguously 
defined bulk melting temperature $ T_m $, found to be 
4510 K using the interatomic potential LCBOPII.
This value is confirmed by simulations for finite disks and
ribbons, which by extrapolation yield $ T_m$'s close to that from 
nucleation theory, and show reversibility to some extent.

Our finding that $ T_m $ is higher (about 250 degrees) than 
for graphite is not likely to be an artifact of LCBOPII. 
As a qualitative explanation we suggest that the increased
stability of graphene is due to a significant positive entropy 
contribution for the solid phase due to the possibility of relatively 
large out-of-plane fluctuations (rippling) at low energy costs, 
as compared to the situation in graphite.

We expect that the melting scenario found for graphene 
applies to any covalent 2D material. This expectation is 
supported by recent melting simulations for 2D MoS$_2$,
showing coexistence of the 2D solid and liquid phase during
melting \cite{SinghMOS2}, as we observed for graphene. 
While measuring $ T_m $ for graphene seems to be a real 
challenge, measurement of $ T_m $ becomes easier for 
2D material with a (much) lower $ T_m $ than that of graphene.

\noindent
$ {\bf Acknowledgements} $.The research leading to these results has received 
funding from the European Union Seventh Framework Programme under grant agreement 
n°604391 Graphene Flagship and from the  Foundation for Fundamental Research on Matter (FOM), which is part of the Netherlands Organisation for Scientific Research (NWO).  We thank Luca Ghiringhelli for useful 
discussions.
\section{APPENDIX: Formation energy of Stone-Wales and 57 defects.}
\begin{figure*}[htb]
\hspace*{-8.cm}a \hspace*{9.cm}b

\includegraphics[width=8.5cm,clip]{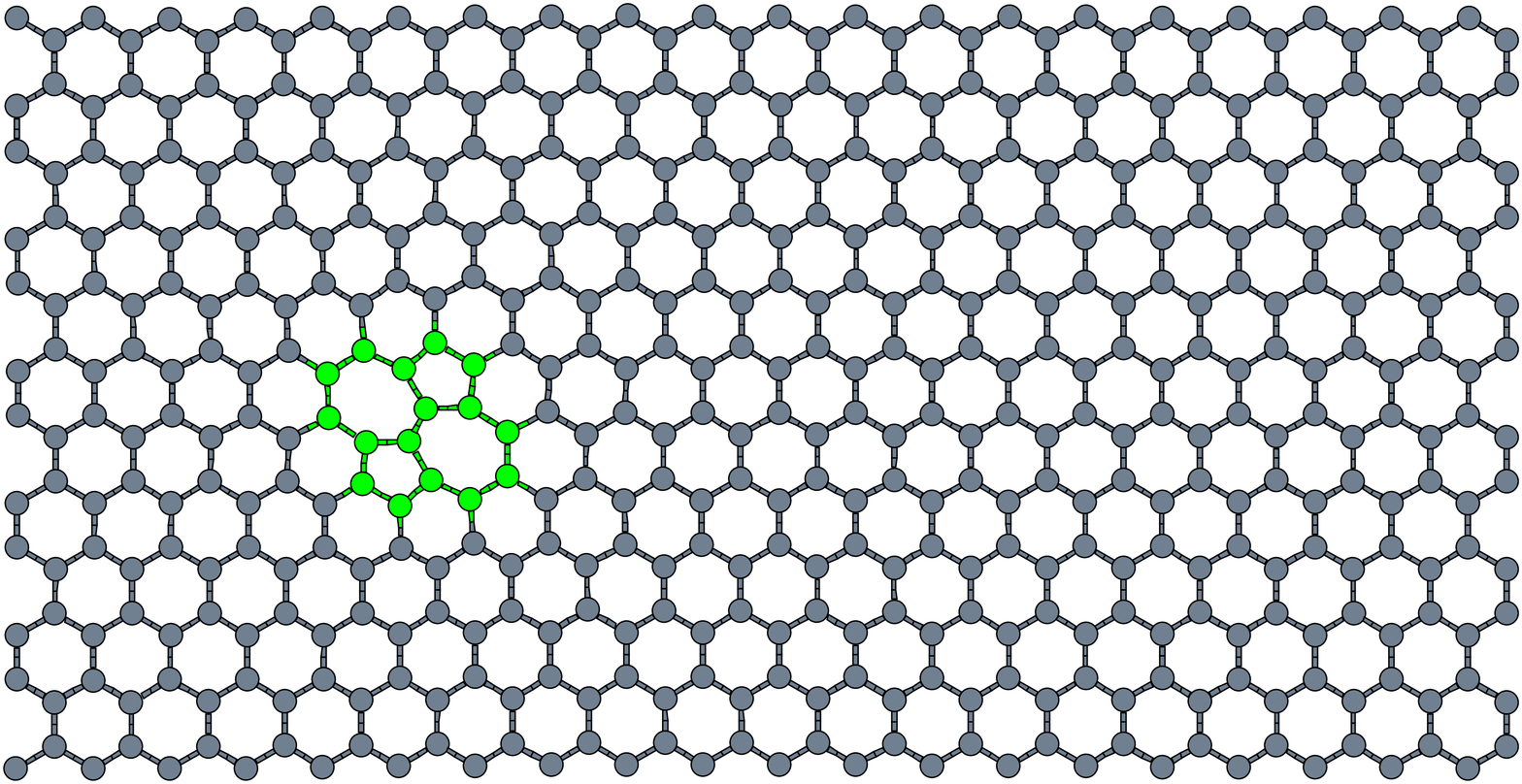} \hspace*{0.5cm}
\includegraphics[width=8.5cm,clip]{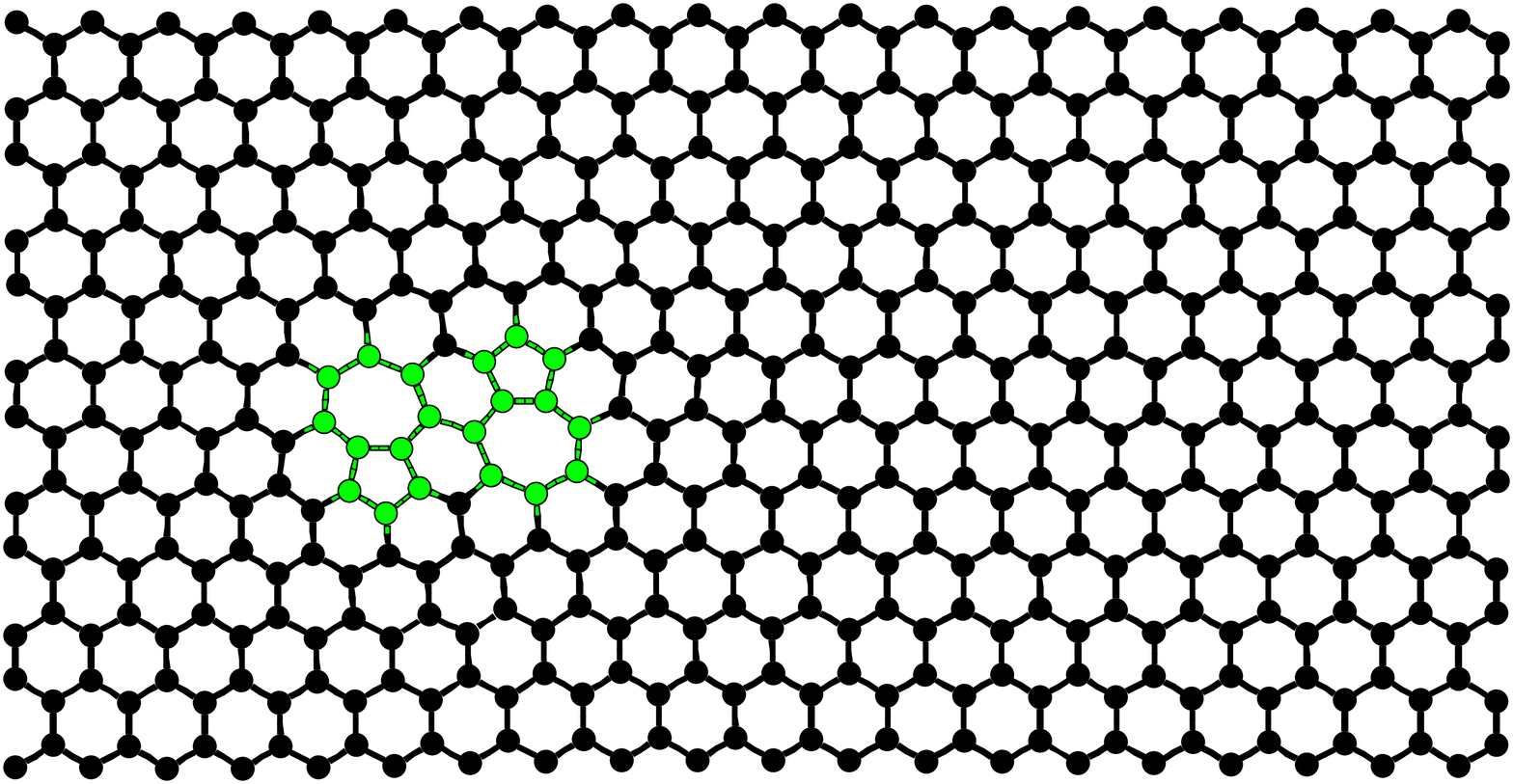}

\hspace*{-8.cm}c \hspace*{9.cm}d

\includegraphics[width=8.5cm,clip]{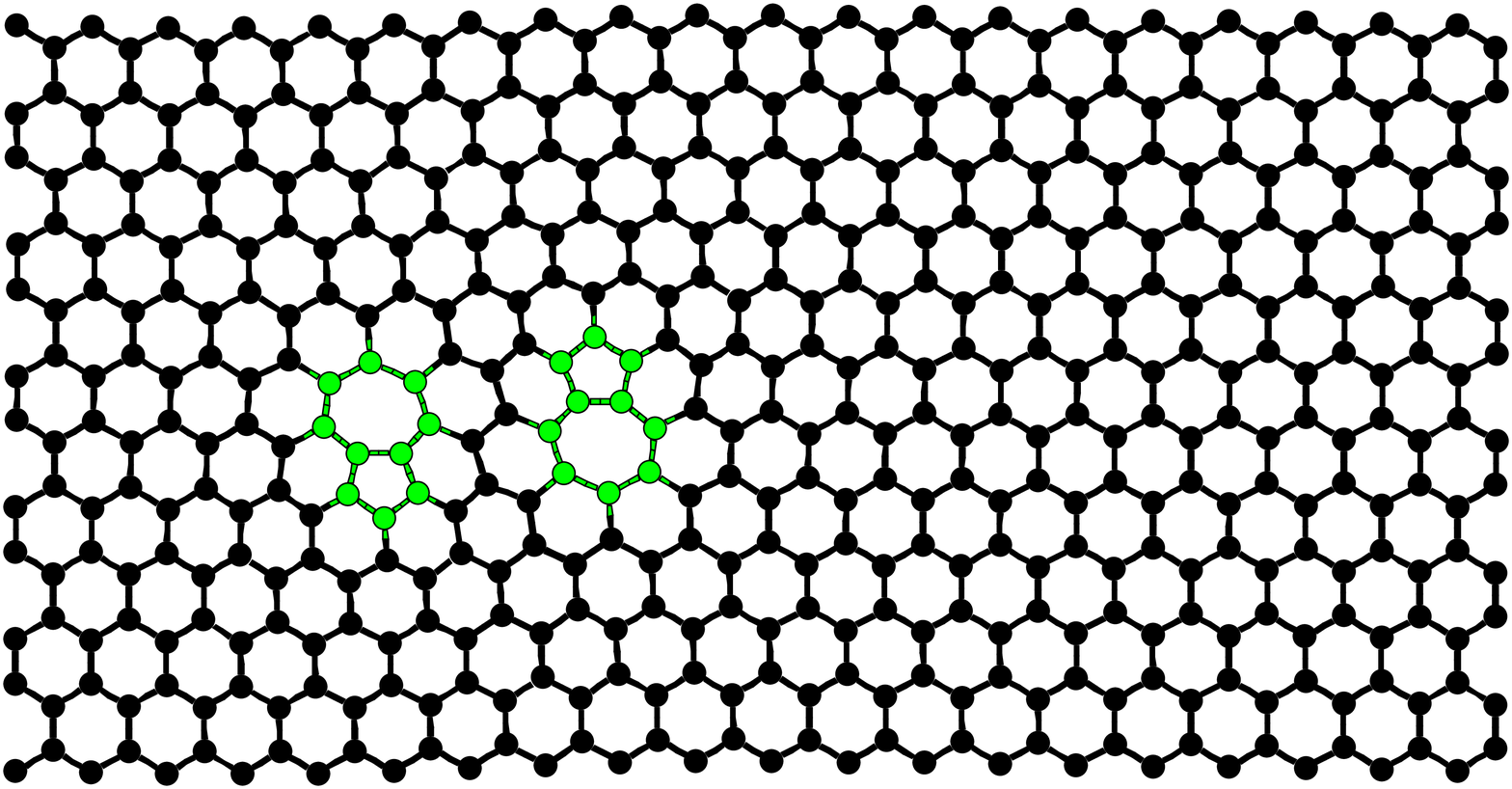} \hspace*{0.5cm}
\includegraphics[width=8.5cm,clip]{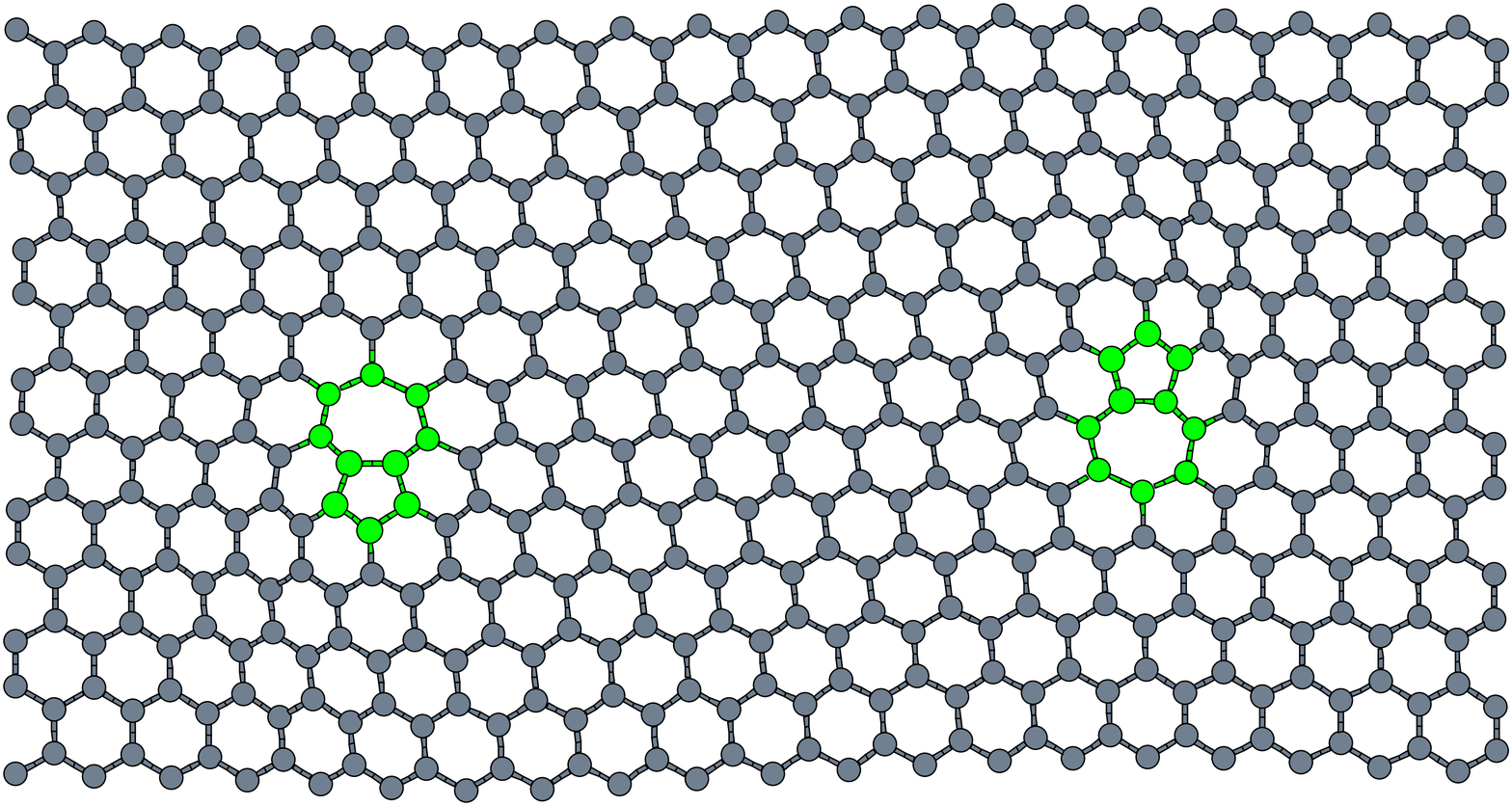}

\vspace*{0.25cm}
\hspace*{-8.cm}e \hspace*{9.cm}f
\vspace*{-0.00cm}

\includegraphics[width=8.25cm,clip]{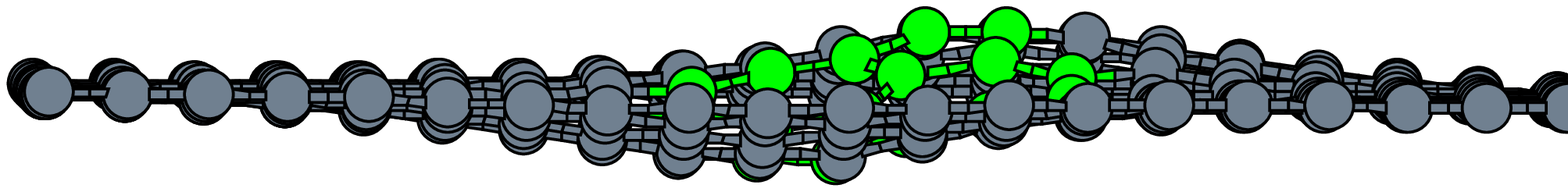} \hspace*{0.5cm}
\includegraphics[width=8.25cm,clip]{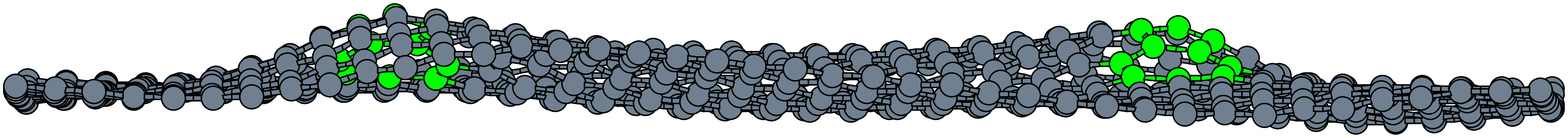}
\caption{\label{conf5775}
Step by step separation of the 5775 SW defect (a) into two 57 defects
with one (b), two (c) and nine (d) hexagonon pairs in between.
Panels e) and f) are side views of the configuration in a) and d), 
showing much larger out-of-plane deformations for the case of 
separated 57 defects, due to strong stress in that case.}
\end{figure*}

A 5775 Stone-Wales (SW) defect can be created in a perfect graphene layer by rotating 
a single bond by 90 degrees, followed by a geometrical relaxation of the system 
(see Fig. \ref{conf5775}a). It consists of two adjacent 57 defects. By subsequently rotating 
another appropriate bond by 90 degrees (and relaxing the system), the 5775 SW 
defects can be split into two 57 defects separated by one pair of hexagons, as
shown in Fig. \ref{conf5775}b. This process can be repeated to separate the 57 defects more 
and more, as shown in Fig. \ref{conf5775}c and d. 

\begin{figure}[htb]
\includegraphics[width=6.cm,clip]{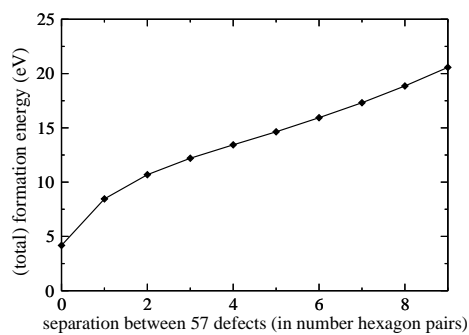}
\caption{\label{E5775} 
Total formation energy as a function of the separation 
between 57 defects according to LCBOPII, calculated for the same sample as
used for Fig.\ref{conf5775}}
\end{figure}

The total formation energy as a function of the separation between the (57) defects
according to LCBOPII is given in Fig. \ref{E5775} and shows that it strongly increases 
with distance. Similar behaviour was previously reported in Ref. \cite{Carlsson}, 
but in that case the 57 defects were created at grain boundaries in a polycrystalline 
sheet. In the latter work it was shown that the formation energies according to LCBOPII
are in very good agreement with those calculated {\it ab initio} within DFT.
The strongly increasing energy implies that the pathway to disorder by the splitting 
of a SW defects into two 57 defects and their subsequent further diffusion, a scenario
in agreement with the KTHNY theory, is very unfavorable. Indeed, we never observed 
separated 57 defects in our simulations.

\end{document}